\documentclass[12pt,preprint]{aastex}
\usepackage{rotating}
\usepackage{graphicx}

\shorttitle{Spectrum of Turbulence in Optically Thick Interstellar Clouds }

\shortauthors{BURKHART ET AL.}
\begin{document}

\title{The Turbulence Power Spectrum in Optically Thick Interstellar Clouds}

\author{Blakesley Burkhart\altaffilmark{1}, A. Lazarian\altaffilmark{1}, V. Ossenkopf\altaffilmark{2}, J. Stutzki\altaffilmark{2}}
\affil{$^1$ {Astronomy Department, University of Wisconsin, Madison, 475 N.  Charter St., WI 53711, USA}}
\affil{$^2$ {Physikalisches Institut der Universit\"{a}t zu K\"{o}ln, Z\"{u}lpicher Strasse 77, 50937 K\"{o}ln, Germany}}

\begin{abstract}
The Fourier power spectrum  is one of the most widely used statistical tools to analyze the nature of magnetohydrodynamic
turbulence in the interstellar medium.   Lazarian \& Pogosyan (2004) predicted that the spectral slope
should saturate to -3 for an optically thick medium and many observations exist in support of their prediction.  However,   
there have not been any numerical studies to-date testing these results.  We analyze the spatial power spectrum
of MHD simulations with a wide range of sonic and Alfv\'enic Mach numbers, which include radiative transfer effects of the $^{13}$CO transition.   
We confirm numerically the predictions of Lazarian \& Pogosyan (2004) that the spectral slope of line intensity maps of an optically thick medium saturates to -3.  Furthermore, 
for very optically thin supersonic CO gas, where the density or CO abundance values are too low to excite emission in all but the densest shock compressed gas, we find that 
the spectral slope is shallower than expected from the column density.
Finally,  we find that mixed optically thin/thick CO gas, which has average optical depths on order of unity, shows mixed behavior:
for super-Alfv\'enic turbulence, the integrated intensity power spectral slopes generally follow the same trend with sonic Mach number as the true column density power spectrum slopes. 
However, for sub-Alfv\'enic turbulence the spectral slopes are steeper with values  near -3 which are similar to the very optically thick regime.

\end{abstract}
 
 \section{Introduction}
\label{intro}

The interstellar medium (ISM) is turbulent on scales ranging from kilo-parsecs to sub-AU
(see Armstrong et al 1995, Elmegreen \& Scalo 2004; Chepurnov \& Lazarian 2010), with
an embedded magnetic field that influences its
dynamics. Magnetohydrodynamic (MHD) turbulence is accepted to be of key importance
for fundamental astrophysical processes, e.g. heat transport,  star
formation, and acceleration of cosmic rays.

Despite the clear importance of MHD turbulence to astrophysics, it is difficult to study.
In light of this, numerical simulations have
tremendously influenced our understanding of the physical
conditions and statistical properties of MHD turbulence (see Vazquez-Semadeni et al. 2000,
Mac Low \& Klessen 2004, Ballesteros-Paredes et al. 2007,
McKee \& Ostriker 2007 and ref. therein). Present codes can
produce simulations that resemble observations in terms of structures and scaling laws, but because
of their limited numerical resolution, they can not reach the observed Reynolds\footnote{Reynolds number ($Re$) is the ratio of the large eddy 
turnover rate $\tau^{-1}_{eddy}=V/L_f$ and the viscous dissipation rate $\tau_{dis}^{-1}=\eta/L^2_f$, where $V$ is the flow velocity, $L_f$ is the injection scale, and $\eta$ is the resistivity of the fluid.} numbers of the ISM (see e.g. McKee 1999, Shu et al. 2004; Lazarian 2009).

In this respect, observational studies of turbulence can test
to what extent the numerical simulations are able to reproduce the conditions in the ISM.  
The turbulence power spectrum, which is a statistical measure
of turbulence that quantifies how much energy resides on a given scale, can be used to compare observations with
both numerical simulations and theoretical predictions.

However, the importance of obtaining the turbulence spectrum
from observations extends beyond testing the accuracy
of numerical simulations.  In general, the shape of the  energy spectrum is determined by a
complex process of non-linear energy transfer and observational studies of the turbulence spectrum
are critical to  determine sinks and sources of astrophysical turbulence.
At large scales $l$, i.e.\ at small wavenumbers $k \sim 1/l$, 
one expects to observe features in the energy spectrum $E(k)dk$ that reflect energy injection.  

In terms of the ISM, 
the injection scale and main energy injectors are still unknown, but it is clear that
turbulence in the Galaxy is driven on large scales (kpc) by supernova, magnetic rotational instability (MRI), galactic
fountains, high-velocity cloud impacts, or some combination of these.  
At small scales one should see the scales corresponding to the 
dissipation of energy.

The hydrodynamic counterpart of MHD turbulence theory is the famous Kolmogorov (1941) theory of turbulence.
The transfer of energy from large scale eddies to smaller scales continues until the cascade reaches eddies that are small enough to dissipate energy over an eddy turnover time. In the absence of compressibility, the hydrodynamic cascade of energy is E$\sim v^2_l/\tau_{casc, l} =const$, where $v_l$ is the velocity at the scale $l$ and the cascading time for the eddies of size $l$ is $\tau_{cask, l}\approx l/v_l$. From this the well known relation $v_l\sim l^{1/3}$ follows.
In terms of the direction-averaged energy spectrum this gives the famous
Kolmogorov scaling $E(k) \sim 4\pi k^2P(k) \sim k^{-5/3}$, where $P(k)$ is the 3D energy spectrum.

The ISM is both turbulent and magnetized, and therefore Alfv\'enic perturbations are vital. 
Contrary to Kolmogorov turbulence, in the presence of a dynamically important magnetic field, eddies become anisotropic along the magnetic field lines while the cascade of energy
proceeds perpendicular to the local magnetic field.  
This observation corresponds to theoretical expectations of the Goldreich \& Sridhar (1995, henceforth GS95) theory of Alfv\'enic turbulence\footnote{The GS95 model was criticized initially as low resolution numerical simulations showed slopes that were shallower than the predicted -5/3.  However, more recent studies
with simulations that are able to resolve the inertial range show that the spectrum converges to -5/3
 (see Beresnyak \& Lazarian 2009, 2010, Beresnyak 2012)}.
For the perpendicular eddies, the original Kolmogorov treatment is applicable resulting in perpendicular motions scaling as $v_l\sim l_{\bot}^{1/3}$, where $l_{\bot}$ 
denotes eddy scales measured perpendicular to the local direction of magnetic field (see Lazarian \& Vishniac 1999, Cho \& Vishniac 2000,
Maron \& Goldreich 2001, Cho, Lazarian \& Vishniac 2002, see Cho et al. 2003 for a review).  
These mixing motions induce Alfv\'enic perturbations that determine the \textit{parallel} size of the magnetized eddies, i.e.,  the  {\it critical balance} condition. 
Critical balance in GS95 is the equality of the eddy turnover time $l_{\bot}/v_l$ and the period of the corresponding Alfv\'en wave $\sim l_{\|}/V_A$, 
where $l_{\|}$ is the parallel eddy scale and $V_A$ is the Alfv\'en velocity. Making use of the earlier expression for $v_l$ one 
obtains $l_{\|}\sim l_{\bot}^{2/3}$, which reflects the
tendency of eddies to become more and more elongated as the energy cascades to smaller scales. The relations for incompressible MHD turbulence carry over
to the Alfvenic and slow modes of compressible MHD (Cho \& Lazarian 2002, 2003, Kowal \& Lazarian 2010). 

In light of the theoretical progress on studies of the power spectrum scaling of turbulence, 
there have been many investigations
over the last ten years which study the density/velocity power spectrum in radio
position-position-velocity (PPV) 
cubes of neutral hydrogen in
the Milky Way Galaxy,  the Magellanic clouds and other galaxies in the context of turbulence
(see Table 1 for a summary of these studies).  This is complex due to the entanglement of density and velocity fluctuations in PPV space.
Attempts to use PPV data cubes to study fluctuations of intensity can be traced back to 
the work
by Crovisier  \& Dickey (1983), Green (1993)
and Stanimirovic et al. (1999). These studies suffered from a lack of theory with which to relate the statistics of PPV fluctuations with the underlying statistics of velocity and density fluctuations.  Thus the interpretation of the measured spectra of intensity fluctuations in channel maps was highly uncertain.

These investigations 
do not only study the 2D column density spectrum, but also 
analyze 
the spectrum with varying channel thickness in the radio cube.  
This is the central idea behind the Velocity Channel Analysis (VCA) developed by Lazarian \& Pogosyan (2000, henceforth LP00) and in subsequent 
papers (Lazarian \& Pogosyan (2004,2006) which is intended to obtain both velocity and density turbulence spectra from the observations. 
The validity of the technique was verified numerically for optically thin gas in Esquivel et al. (2003) and at much higher precision in 
Chepurnov \& Lazarian (2009).   
However, no systematic numerical study has been performed so far for the study of the intensity power spectrum in the presence of a self-absorbing media, although several previous studies have 
discussed the importance of radiative transfer effects on the statistics of turbulence (e.g, Padoan et al. 2003; Shetty et al. 2011).

The contribution
of the velocity fluctuations as outlined by LP00 may depend on whether the
images of the eddies under study fit within a velocity slice (a ``thick slice'')
or if their velocity extent is larger than the slice thickness (a ``thin slice'').
The spectra of fluctuations that correspond to thin and thick
slices are different and varying the thickness
(i.e. effective channel width) of slices provides the possibility to disentangle
the statistics of underlying velocities and densities
in the turbulent volume.  More recently, a technique that analyzes the density and velocity spectrum of turbulence by taking the spectrum along the velocity axis, the Velocity Coordinate Spectrum (VCS), 
has also been developed and is complementary to the VCA technique (see Lazarian \& Pogosyan 2008; Lazarian 2009; Chepurnov et al. 2010).
Table 1 summarizes some of the variety of objects
to which VCA has been applied.

How do these \textit{observed} spectral slopes relate to predictions given by theory and numerics?
The VCA/VCS techniques were developed from purely analytical considerations, and hence the spectral slope obtained from them
can be related back to turbulence parameters such as the compressibility (due to shocks),  the injection scale, the temperature of the medium and 
energy contained in the turbulence. 
Furthermore, it is encouraging that the observed spectral indexes correspond
to what is expected from simulations (see Beresnyak, Lazarian \&
Cho 2005; Kowal, Lazarian \& Beresnyak 2007;  Burkhart et al. 2010), which show that, as the sonic Mach number increases, 
the density spectrum becomes increasingly flatter  while the spectrum of velocity gets steeper.
For incompressible hydrodynamic or super-Alfv\'enic turbulence, high resolution simulations show very good agreement with the $k^{-5/3}$ slope (Beresnyak 2012). In nearly incompressible
motions with a relatively strong magnetic field, the spectrum of density scales similarly to
the pressure as  $k^{-7/3}$ (Biskamp 2003; Kowal, Lazarian \& Beresnyak 2007; Burkhart et al. 2010). These theoretical results are summarized in Table 2.\footnote{We would like to stress that, for incompressible turbulence, the Kolmogorov power spectrum in three dimensions (3D)
is $k^{-11/3}$, in 2D it is $k^{-8/3}$, and 1D $k^{-5/3}$ for the same energy spectrum E(k) and in this paper we use the -11/3 slope for Kolmogorov.}

These theoretical and numerical predictions are almost always done under the assumption of an optically thin medium.
However,  Lazarian \& Pogosyan (2004, henceforth LP04) predict that absorption
can induce a universal spectrum â$k^{-3}$.  Inspection of Table 1 shows that this
prediction has already garnered observational evidence
although it has not yet been tested numerically.
In this paper, we use numerical simulations with radiative transfer effects simulating the $^{13}$CO $J=2-1$ transition.  
We vary our radiative transfer parameter space, including CO abundance ($^{13}$CO/H$_2$) and average density ($n$),  to sample optically thin, optically thick, and mixed optically thin/thick lines.
In this way, we can test how well the recovered 2D density spectra match the theoretical predictions of the above mentioned 
investigations. 

This paper studies the effects of self-absorption on the recovery of the underlying density spectrum from integrated intensity maps. For this purpose, we do not use the idealized analytical model of radiation transfer adopted in LP04 but a radiative transfer code that
is described in Ossenkopf 2002 and Burkhart et al. 2013b.    We compare our results with the theoretical predictions in LP04 and investigate the conditions in which the integrated intensity maps with varying optical depth reflect the underlying density spectrum of turbulence.

This paper is organized as follows: 
In \S~\ref{sec:Sims} we describe the simulations and outline the main points of the radiative transfer algorithm.
In \S~\ref{sec:spect} we present the analysis of the 2D power spectrum of our simulations including optical depth effects.
Finally, in  \S~\ref{sec:dis} we discuss our results followed by the conclusions in \S~\ref{sec:con}.

\section{Data and Method}
\label{sec:Sims}
We generate  3D numerical simulations of isothermal compressible (MHD)
turbulence by using the the Cho \& Lazarian  (2003) MHD code and varying the input
values for the sonic and Alfv\'enic Mach number. We describe the simulations here as they are presented in code units, however
the isothermal simulations can be scaled to physical units easily as they are scale-free (see the appendix of Hill et al. 2008 for more information on scaling).
However, once radiative transfer is introduced, the simulations are no longer scale free. 

Turbulence is driven with large-scale
solenoidal forcing.  The magnetic field consists of the uniform background field and a
fluctuating field: ${\bf B}= {\bf B}_\mathrm{ext} + {\bf b}$. Initially ${\bf b}=0$.
For more details on the numerical scheme see Cho et al. (2002), Cho \& Lazarian (2003), Kowal, Lazarian \& Beresnyak (2007) and Burkhart et al. (2009).

We divided our models into two groups corresponding to
sub-Alfv\'enic ($B_\mathrm{ext}=1.0$) and
super-Alfv\'enic ($B_\mathrm{ext}=0.1$) turbulence.
For each group we compute several models with different values of
gas pressure, which is our control parameter that sets both the sound speed and the sonic Mach number (see Table \ref{fig:radtab}, second column).
We run compressible MHD turbulent models, with 512$^3$ resolution,
for $t \sim 5$ crossing times, to guarantee full development of the energy cascade.  The models are listed and described in Table \ref{fig:radtab}. 

After we generate the simulations, including the full 3D density and velocity
cubes, we apply the SimLine-3D radiative transfer algorithm (Ossenkopf 2002) for
the $^{13}$CO 2-1 transition. The code computes the local excitation of
molecules from collisions with the surrounding gas and from the radiative
excitation at the frequencies of the molecular transitions through line and
continuum radiation from the environment. Instead of an exact description of
the mutual dependence of the radiative excitation at each point in
a cloud on the excitation at every other point, the code uses
two approximations to describe the radiative interaction. First it computes
a local radiative interaction volume limited by the velocity gradients (LVG). Cells
with line-of-sight velocities different by more than the thermal line width
cannot contribute to the excitation of the considered molecule, i.e. the
interaction length is limited by the ratio of the thermal line width to the
velocity gradient. This is a particularly useful approximation for supersonic
turbulence. For the interaction of more remote points that accidentally
have the same line of sight-velocity, the second approximation applies, using
the average isotropic radiation field in the cube for the excitation at
every individual point. This best covers all cases with isotropic
structure, i.e. turbulence simulations which are not
dominated by only a few large-scale structures.
The driving of turbulence is done on large scales ($k\approx 2$) and most structure (i.e. total power), including our inertial range, is at scales smaller than this. 
The error contributing to the power spectrum will be largest for large structures with internal radiative pumping as  the detailed shape of the radiative interaction volume 
is ignored and instead replaced by an ellipsoid determined from the orthogonal velocity gradients. For all our cases, the approximations should be very good, 
so that we expect an accuracy of better than 10\% for supersonic turbulence and 20\% for subsonic turbulence. 
This level of accuracy is sufficient for
comparison with observational data as drifts in the receiver system and the atmosphere
and the resulting temporal variation of the calibration parameters typically also
provide calibration errors of that magnitude.

For our tests we choose a total cube size of 5pc and a gas temperature of 10K.
The cube is observed at a distance of 450pc with a beam FWHM of 18'' and a
velocity resolution of $0.05\rm{kms}^{-1}$. We vary both the density scaling factor
(in units of $\rm{cm}^{-3}$, denoted with the symbol $n$) and the molecular abundance
($^{13}$CO/H$_2$, denoted with the symbol $x_{co}$) in order to change
excitation and optical depth $\tau$. To represent the
typical parameters of a molecular cloud we choose standard values for density,
$n=$275 $\rm{cm}^{-3}$, and abundance, $x_{co} = 1.5\times10^{-6}$ (what we will refer to as the normalized or norm parameter setup), 
providing an slightly optically thick $^{13}$CO 2-1 line at the highest densities with line
center optical depths of a few ($\tau$(norm) in Table \ref{fig:radtab} in
column 4). To investigate the impact of line saturation and subthermal
excitation we vary density ($n$) and abundance ($x_{co}$) by
factors of 30 to larger and smaller values compared to the $\tau$(norm) values.
We list all the parameters of the complete model set in Table~\ref{fig:radtab}.

\section{Results}
\label{sec:spect}

\begin{figure*}[tbhp]
\centering
\includegraphics[scale=.5]{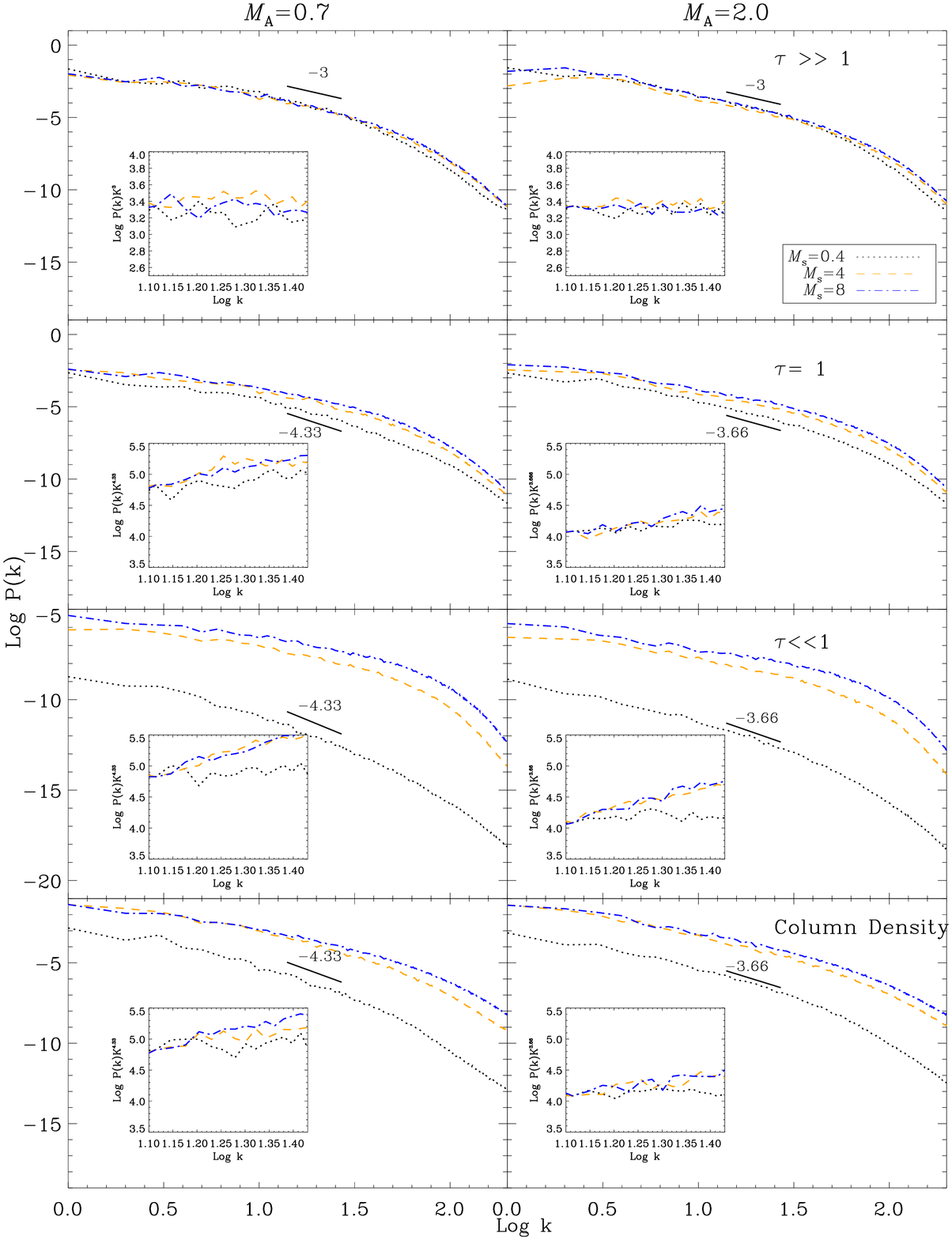}
\caption{Energy spectrum vs. wavenumber on a log-log scale for models 1 ,3, 5, 6, 8,10 from Table 3 for varying density (Column 4-6 of Table 3). 
The top row represents the most optically thick cases (with $n8250$), the next two rows down are for the norm case and $n 9$ cases, respectively,
and the very bottom row shows the column density power spectrum with no effects of radiative transfer.  The sub-Alfv\'enic cases are organized on the left panels
and the super-Alfv\'enic cases are organized on the right panels. Thick solid lines represent the theoretically expected 3D power spectrum given in Table 2 (i.e. the Kolmogorov slope is -11/3).
We find almost identical spectral trends for cases of varying abundances (Column 7-8 of Table 3) and hence only make the plot for the density parameter space.
We show  inserts with the power spectrum normalized both on an absolute scale to the amplitude at log $k=1.1$ and to the theoretical slopes expected from Table 2.
The k range plotted on the x-axis of the inserts is identical to the range we take for the fitted slopes.  }

\label{fig:powerspectrum}
\end{figure*}

The energy spectrum vs. wavenumber for models 1, 3, 5, 6, 8, 10 from Table 3 for varying density (Column 5-7 of Table 3) is shown
in Figure \ref{fig:powerspectrum}. 
The left column shows the sub-, the right column the super-Alfv\'enic case. 
We find almost identical "by eye' power spectral trends for cases of varying abundances 
(i.e. models in column 8-9 of Table 3) and hence only make the plot for the varying density parameter space (i.e. holding abundance constant). 
Additionally, we also do not find significant differences between the power spectrum of different lines-of-sight relative to the mean magnetic field and only show here
a line-of-sight taken perpendicular to the mean field.

\begin{figure*}[tbhp]
\centering
\includegraphics[scale=.6]{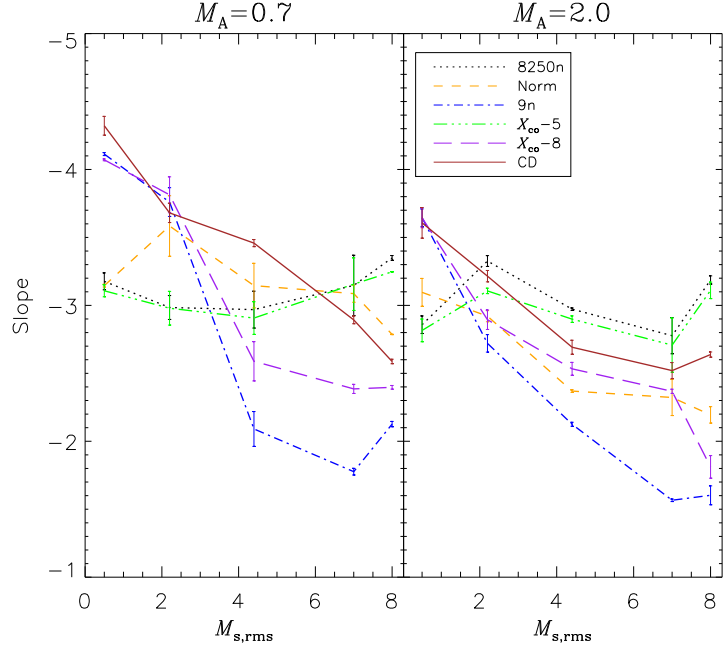}
\caption{Fitted slopes  vs. sonic Mach number for the entire parameter space listed in Table 3. Error bars are 
created by moving the fitted slope range from $k=13-24$ to $k=13-26$ and taking the standard deviation of the slopes in these ranges.}
\label{fig:slopes}
\end{figure*}

The left and right top panels represent 
the most optically thick cases (with $n8250$), the next two panels down are for the norm case and $n 9$ cases, respectively,
and the very bottom panels show the column density power spectrum with no effects of radiative transfer. 
Thick solid lines represent the theoretically expected 3D power spectrum given in Table 2. 
The k range of this line represents the range that we fit the spectral slope in Figure \ref{fig:slopes}.  The sub-panels in each plot show the power spectrum
normalized both by the amplitude at log $k$=1.1 and the theoretically expected slope:  for the optically thick power spectrum (top two panels) the normalization slope is  $k^3$, for super-Alfv\'enic turbulence it is $k^{3.66}$ and for sub-Alfv\'enic turbulence it is $k^{4.33}$.

The bottom two rows of  Figure \ref{fig:powerspectrum} (for the true column density and optically thin simulations) show that, for both
sub- and super-Alfv\'enic turbulence, the incompressible case (black dotted line) follows closely the 
theoretically expected slope.  The yellow and blue dashed lines, which are for compressible simulations, show slopes
that are shallower than the incompressible simulations and the Kolmogorov slope, which again converges well with the expectations outlined
in Table 2.   For the $\tau \approx 1$ cases (second row down from the top) we see similar trends, however it appears that all cases have slopes
slightly shallower than the slopes of the column density spectra. 

Inspection of the top row reveals that, when the optical depth is very large, there is no longer any clear distinction between
simulations with different sonic and Alfv\'enic Mach numbers.  It appears that, despite any difference in compressibility or magnetization the spectral slope converges around
$\approx$ -3.

We perform a fit to the slopes in the inertial wavenumber range of $k=13-26$ (shown with the solid black line in Figure 1) and $k=13-24$, averaging the slopes in these ranges 
and creating error bars using the standard deviation. 
We plot the slope vs. sonic Mach in Figure \ref{fig:slopes} for our full parameter space listed in Table 2, including varying density and abundance.
The column density (red-brown solid lines)  follow closely what is expected from Table 3 for all sonic and Alfv\'enic Mach numbers. 
Namely, we find 
that the incompressible slopes are $\approx$ -3.66 and -4.33 for super- and sub-Alfv\'enic turbulence, respectively, and then become increasingly shallow as the sonic Mach number increases.  
Similar to the findings of Burkhart et al. 2010, 
the slopes approach $\approx -2.5$ for high sonic Mach numbers regardless of the Alfv\'enic Mach number.

The very optically thick cases (dotted black and dash dotted green lines) are saturated around -3 regardless of the Sonic or Alfv\'enic Mach number with some variations.  
These variations at most extend up to -3.3 for the steepest cases
 and to -2.8 in the shallowest cases.  
On might expect that the use of high resolution simulations with radiative transfer might show slopes converging even closer to -3.

Interestingly, the mixed optically thin/thick case or the norm case (yellow dashed lines)  which has optical depth on order of unity, 
shows a trend with the slope that is dependent on the Alfv\'enic Mach number.  For super-Alfv\'enic  turbulence, 
the  norm case follows the column density slopes and the slopes are much shallower than compared with the sub-Alfv\'enic counterpart (a difference in the slope of 0.6 for the $M_s\approx 8$ simulations).  However, for sub-Alfv\'enic  turbulence, the norm case behaves as though
it were generally optically thick, and the slopes remain somewhat more around -3, although the $M_s \approx 2.0$, $M_A \approx 0.7$ case shows a larger variation in its slope of $\approx -3.5 \pm 3$.  
Inspection of Table 3 reveals the 
optical depths for the norm case are not substantially different between models with high and low magnetic field (all hover around $\tau \approx 3-4$).

Finally, we also plot the slopes for the very optically thin cases where the density or abundances are very low (dotted dashed blue lines and thick dashed purple lines) in Figure 2.  
 For the incompressible (subsonic) cases, the values of the slopes converge well
with the true column density values, especially for super-Alfv\'enic turbulence (slopes of around -3.6). 
However, as the sonic Mach number
increases, the slopes for these models become shallower than the column density slopes. 
The super-Alfv\'enic slopes are slightly shallower than the sub-Alfv\'enic slopes.
We conclude that, for supersonic turbulent CO integrated intensity maps with very low density/abundance values,  the power of the total intensity map is increasingly on smaller scales as compared with the true column density. 
 This is due to the fact that, for these cases, excitation of the $^{13}$CO line is limited due to the density/abundance being too low to excite regions
that are not in the highest density areas i.e. shock compressed regions.  Thus, a spectrum that is shallower than the column density power spectrum is obtained.

\section{Discussion}
\label{sec:dis}

\subsection{Optical Depth Studies on the Turbulence Power Spectrum}
Our study supports the theoretical conclusion in LP04 that the spectrum of
line intensity fluctuations in the case of optically thick lines tends to the universal spectral slope of -3, which does not reflect the actual
underlying turbulence. In general, we also found that slopes shallower than the predictions of Table 2 can result from higher sonic mach numbers, lower optical depths and, to a lesser extent,  super-Alfv\'enic turbulence.

We varied both density and CO abundance in our simulations to vary the optical depth in order to investigate its effects on the power spectrum.
Our standard abundance
represents the abundance in a fully molecular material. The transition
region investigated by Clark et al. 2012 represents an intermediate regime
towards our low-abundance case. For studies investigating the Fourier power spectrum of line intensity maps, we have shown that the main effect
to be taken into account is the change of the
optical depth, rather than the specific value of density or abundance.

However, one might speculate that the spatial variations of the CO abundance (Glover et al. 2011) can be reflected in the power spectra.
The abundance change will
occur on the larger scales, i.e.  k $\sim 1$, which is
not in the inertial range that we consider here.
Therefore we can treat the abundance change in those simulations
as a global change of the inertial power spectrum, similar to the global
abundance scalings that we perform manually when going from $x_{co}-8$
to norm and to $x_{co}-5$. In that sense our experiment can be directly
used to study the effect of different abundance scalings in the process of
molecular cloud formation. When interpreting our results e.g. directly
in terms of $^{13}$CO observations, we will therefore find a transition
between the optically thin case and the norm case, i.e. we only expect a
small change of the power spectrum slope in the process of the
molecular cloud formation. In contrast, for the main $^{12}$CO isotope
the molecular cloud formation process rather represents the
transition from the norm case to the optically very thick cases, i.e.
we predict a change of the power spectrum slope to -3 in the
molecular clouds.

We note that even in the case of optically thick lines, the power spectrum of channel maps may still
reflect the velocity and density spectra for sufficiently large separations (see the corresponding criterion in LP04). However, the separation of density and velocity contributions may be more difficult using
the VCA technique.
For a steep density spectrum, which is expected for  subsonic turbulence,  we can recover the velocity spectral index. For supersonic turbulence that produces a shallow density
spectrum, the recovery for the velocity spectral index from thin channels faces the problem that we do not know the actual density spectral index.

Thus, in order to know which regime we are in, we must know the sonic Mach number.
The issues of the Mach numbers of turbulent clouds can be resolved with the use of other techniques (see Kowal et al. 2007; Burkhart et al. 2009; Burkhart \& Lazarian 2012) or 
from independent velocity and temperature measurements (see Burkhart et al. 2010; Kainulainen \& Tan 2012).  In this case, if one is in the subsonic regime, one can get the velocity spectrum directly. 
In the case of supersonic turbulence one may compare the fit of the velocity-density spectral to numerical simulations. 
Alternatively, the density spectrum can be obtained independently,
e.g. from column density and dust extinction maps  (see Lazarian 2009). This calls for  multi-frequency and multi-instrument studies of turbulence. At the same time, comparing the spectra
 of density from dust and of total line emission can gauge the effects of optical absorption.\footnote{As both studies of Padoan et al. 2006 and Chepurnov et al. 2009
resulted in the spectral index of line emissivity close to -3, it is important to compare these results with those from dust emission.}

We note that for both VCA and VCS the thermal broadening is important. However, if we use sufficiently heavy species, the thermal broadening is reduced and therefore both subsonic and supersonic turbulence can be studied.

In addition, our study confirms that also for optically thin species, the spectral slope depends on the line excitation. When density and abundances are low, we are under-sampling most of the gas, 
causing emission to be concentrated in the shock dominated regions.  This has the effect of shallowing out the spectral slope in the case of supersonic turbulence.  Subsonic turbulence is hardly affected.

\subsection{Differences in the Spectral Slope Between Sub-Alfv\'enic  and Super-Alfv\'enic Turbulence with Varying Optical Depth}

While previous publications 
have predicted differences in the power spectral slopes between hydrodynamic/super-Alfv\'enic and highly magnetized turbulence in the incompressible limit with no radiative transfer effects (See Table 2) and in the case of fully optically thick lines (LP04), no predictions exist for determine the influence the magnetic field has on the slopes with optical depths around unity or less.   In this work we present the first measured differences between the slopes of
 these magnetic regimes with varying optical depth.

We find a significant difference in the regime of moderate optical
depths (norm case), where the synthetic $^{13}$CO map shows a steeper
spectrum for the supersonic, sub-Alfv\'enic turbulence compared to
the super-Alfv\'enic case. The spectrum shows more power on small
scales similar to the transition to optically thick lines. This
indicates a higher radiative excitation of the synthetic $^{13}$CO 2-1
transition in the sub-Alfv\'enic case, increasing the optical depth
of the line. It can be related to the "radiative interaction volume",
i.e. the size of the volume around any point that has approximately
the same line-of-sight velocity. For subsonic turbulence, this volume
is given by the whole cube size, but for supersonic turbulence it
is determined by the shape and size of the turbulent eddies. Due to
the elongation of the eddies in sub-Alfv\'enic turbulence, their volume
is increased as compared with their super-Alfv\'enic and hydrodynamic
counterparts. Additionally, the stronger magnetic field decreases the compression of the media
and mitigates the shock formation, which also increases the radiative interaction volume.
Thus we obtain a higher radiative pumping as a result
of the increase in eddy volume, finally leading to a power spectrum
that falls between the column density scaling and the optically thick
case.  In light of these effects, future work should further investigate the statistics of the velocity field (i.e. line widths)
on the radiative transfer parameter space and the turbulence parameter space.

\section{Conclusions}
\label{sec:con}

We analyze the observable 2D power spectrum of integrated intensity $^{13}$CO maps created from 3D MHD simulations with a range of sonic and Alfv\'enic Mach numbers and optical depths.   

\begin{itemize}
\item  We confirm numerically the predictions of LP04 that the line emission spectral slope of a optically thick medium saturates to -3.
\item  For very optically thin  supersonic CO gas, where the density/abundance values are too low to excite emission in all but the densest shock compressed gas, we find that the spectral slope is shallower than the expectations for column density.
\item We find that mixed optically thin/thick CO gas, which has optical depths on order of unity, shows mixed behavior:
for Super-Alfv\'enic turbulence, the spectral slopes follow the column density slope with $M_s$ trends, while for sub-Alf\'enic
turbulence, the spectral slope is around -3, similarly to the very optically thick regime.  

\end{itemize}

\acknowledgments
B.B. acknowledges support from the NSF Graduate Research Fellowship and the NASA Wisconsin Space Grant
Institution. A.L. thanks both NSF AST 0808118 and the Center for Magnetic Self-Organization in Astrophysical and Laboratory Plasmas for financial support.
All authors acknowledge support through grant SFB956/DFG and by the Deutsche Forschungsgemeinschaft, DFG, project number Os 177/2-1.
Part of this work was completed  during the stay of A.L. as
Alexander-von-Humboldt-Preistr\"ager at the Ruhr-University Bochum and the University of Cologne.

\begin{sidewaystable}
\begin{center}
\begin{tabular}{cccccccc}
\hline\hline
 & tracer  &object & type &  $P^{thin}_{PPV}$  & $P^{thick}_{PPV}$  & $\tau$& Reference \\
\tableline
1 & HI & Anti-center& Galactic& $K^{-2.7}$ & N/A & Thin&  Green (1993); Lazarian \& Pogosyan (2006) \\
2 &HI& Towards CygA & Galactic &  $K^{-2.7}$& $K^{-2.8}$& Thin& Deshpande et al. (2000) \\
3 & HI & SMC&  extragalactic & $ K^{-2.7}$ &  $K^{-3.4}$ & Thin & Stanimirovic \& Lazarian (2001); Burkhart et al. 2010 \\
4 & HI & Center & Galactic & $K^{-3}$&  $K^{-3}$&  Thick&  Dickey et al. (2001); Lazarian \& Pogosyan (2004) \\
5 & HI  & B. Mag. & Galactic & $K^{-2.6} $& $K^{-3.4}$& Thin&  Muller et al. (2004) \\
6 &HI & Arm& Galactic &  $K^{-3}$&  $K^{-3}$&  Thick&  Khalil et al. (2006); Lazarian (2006)\\
7 &HI &  DDO 210& extragalactic & $K^{-3}$& $K^{-3}$& Thick& Lazarian (2006); Begum et al. (2006)\\
8 &$^{12}$CO& L1512& Galactic & N/A &$K^{-2.8}$& Thick& Stutzki et al. (1998); Dickey et al. (2001) \\
9 & $^{13}$CO & L1512 & Galactic &N/A& $K^{-2.8}$ & Thick & Stutzki et al. (1998); Begum et al. (2006) \\
10 & $^{13}$CO& Perseus & Galactic &$K^{-2.7}$ &  $K^{-3}$ & Thick& Sun et al. (2006) \\
11 & $^{13}$CO & Perseus & Galactic &$K^{-2.6}$ &  $K^{-3}$ & Thick& Padoan et al. (2006) \\
12 & C$^{18}$O & L1551 & Galactic  &$ K^{-2.7}$ & $K^{-2.8}$ & Thin & Swift (2006) \\

\hline\hline
\end{tabular}
\caption{Selected VCA results and references based on different atomic and molecular tracers and objects. $P^{thin}_{PPV}$ corresponds to thin slices of the PPV cubes, which are dominated by velocity fluctuations
while $P^{thick}_{PPV}$ corresponds to thick slices of the PPV cube, which are dominated by density fluctuations.  $\tau$ gives the optical depth nature of data.  
\label{fig:vcares}}
\end{center}
\end{sidewaystable}

\begin{sidewaystable}
\begin{tabular}{cccc}
\hline\hline
Magnetic Nature & Type & Line Intensity Spectrum  & Reference \\
\tableline
$M_A < 1$ & incompressible & $\approx k^{-13/3}$& Biskamp 2003; Kowal et al. 2007\\
$M_A  > 1$ & incompressible& $\approx k^{-11/3}$&  GS95; Lithwick \& Goldreich 2001;  Cho \& Lazarian (2002,2003) \\
          & Compressible & shallower than $k^{-11/3}$& Beresnyak, Lazarian \& Cho 2005; Kowal, Lazarian \& Beresnyak 2007\\
          & optically thick& $\approx k^{-3}$ & Lazarian \& Pogosyan (2004) \\
\hline\hline
\end{tabular}
\caption{Power spectra slopes of turbulence for different environments.  Corresponding references to theoretical and numerical 
are in the far right column. }
\label{tab:slopes}
\end{sidewaystable}

\begin{table*}
\begin{center}
\begin{tabular}{ccccccccc}
\hline\hline
Run & $\approx {\cal M}_s$ & $\approx{\cal M}_A$ & N& $\tau$(norm)& $\tau (n$8250)& $\tau (n$9) &$\tau (x_{co}$-5)&$\tau (x_{co}$-8) \\
\tableline
1 &0.4&0.7&512&3.6&71&0.13&37&0.18\\
2&2&0.7&512&5 &101 & 0.01&104 & 0.12 \\
3&4&0.7&512&3&101&0.13&85&0.17\\
4&7&0.7&512&4.8&72&0.14&99&0.19\\
5&8&0.7&512&3.2&65&0.007&87&0.1\\
6&0.4&2&512&3.42&67&0.12&104&0.16\\
7&2&2&512&4.2&86 & 0.01&84&0.01\\
8&4&2&512&3.5&103&0.07&75&0.1\\
9&7&2&512&2.9&63&0.1&64&0.14\\
10&8&2&512&3.47&81&0.09&70&0.3\\
\hline\hline
\end{tabular}
\caption{Description of the ten simulations used for this study, with model number given in column one. Columns two and three show the average sonic and Alfv\'enic Mach numbers. Column four shows the numerical resolution (N) of the simulations.  In columns five through nine we show the map-averaged line center optical depths for the different
values of our radiative transfer parameter space.  
For our radiative transfer parameter space we  vary both the number density scaling factor
(in units of $\rm{cm}^{-3}$, denoted with the symbol $n$) and the molecular abundance
($^{13}$CO/H$_2$, denoted with the symbol $x_{co}$) in order to change
excitation and optical depth $\tau$. To represent the
typical parameters of a molecular cloud we choose standard values for density,
$n=$275 $\rm{cm}^{-3}$, and abundance, $x_{co} = 1.5\times10^{-6}$ (what we will refer to as the normalized or norm parameter setup shown in column five).
We vary density and abundance  by values that are
factors of 30  larger and smaller as compared to the $\tau$(norm) values.
Thus column six, denoted by $n8250$, has parameters: $n=$8250 $\rm{cm}^{-3}$, and abundance, $x_{co} = 1.5\times10^{-6}$.
Column seven, denoted by $n9$, has parameters: $n=$9 $\rm{cm}^{-3}$, and abundance, $x_{co} = 1.5\times10^{-6}$.
Column eight, denote by $x_{co}-5$, has parameters: $n=$275 $\rm{cm}^{-3}$, and abundance, $x_{co} = 4.5\times10^{-5}$.
Column nine, denoted by $x_{co}-8$, has parameters: $n=$275 $\rm{cm}^{-3}$, and abundance, $x_{co} = 5\times10^{-8}$. 
\label{fig:radtab}}
\end{center}
\end{table*}

\end{document}